\documentclass[preprint,showpacs,prb]{revtex4} 
\usepackage{amsbsy,amssymb}                     % Math packages
\usepackage{graphicx}

\begin{document}

%\begin{frontmatter}

%\input{symbols}
\title{
Theory of the Magnetic Resonance for the High-$T_C$ 
Copper-Oxide Superconductors}
\author{M. Azzouz}
\email{mazzouz@laurentian.ca}
\affiliation{Department of Physics, Laurentian University,
935 Ramsey Lake Road,
Sudbury, ON P3E 2C6, Canada.}
\date{17 March 2014}

\begin{abstract}
The magnetic response expected
from a state characterized by rotating 
antiferromagnetism in a neutron-scattering experiment
is calculated. We predict the occurrence of a peak
at the frequency of the rotation of the rotating antiferromagnetic 
order parameter.
The doping dependence of this frequency is very similar
to that of the frequency of the magnetic resonance
observed in the neutron-scattering
experiments for the hole-doped high-$T_C$ cuprates. 
This leads us to propose the rotating
antiferromagnetism as a possible mechanism for this magnetic resonance.
We conclude that while the magnitude of the rotating antiferromagnetic order
parameter was previously proposed to be responsible for the pseudogap 
and the unusual thermodynamic 
and transport properties, the phase of the 
rotating order parameter is proposed here to be responsible for the unusual
magnetic properties of the high-$T_C$ copper-oxide superconductors.
\end{abstract}

\maketitle

\section{Introduction}

The bulk of the  
experimental data collected so far on the magnetic properties of the 
hole-doped high-$T_C$ cuprate superconductors
can be classified using three main key features or behaviors:
i) The unusual zero momentum (${\bf q}=0$) static antiferromagnetic order 
first discovered by Fauqu\'e {\it et al.}\cite{fauque2006} in the hole-underdoped
YBa$_2$Cu$_3$O$_{6+x}$ system. Subsequently, this unusual order was also 
observed in the same regime for the single-layer 
cuprate HgBa$_2$CuO$_{4+\delta}$.\cite{li2008}
Interestingly, this order develops exactly below
the doping-dependent pseudogap\cite{timusk1998} (PG) 
temperature, indicating the existence
of a connection between 
the PG and this order.
ii) Second, the hour-glass magnetic spectrum shape\cite{fujita2012} 
in general or the magnetic resonance\cite{rossat-mignod1991} in particular, 
which characterizes the magnetic excitation energies 
as a function of momentum for the hole-doped
materials. Contrary to the zero-momentum order,
the hour-glass spectrum highlights clearly 
the importance of the usual antiferromagnetic correlations 
because this spectrum is centered around the
antiferromagnetic momentum ${\bf Q}_{AF}=(\pi,\pi)$, where the 
resonance has been observed for most hole-doped high-$T_C$ cuprates.
iii) And third, magnetic excitations were seen 
within the PG phase by Li {\it et al.}\cite{li2010} in the HgBa$_2$CuO$_{4+\delta}$
system. This material is also characterized by the unusual static zero-momentum  
antiferromagnetism.\cite{li2008} These excitations develop in the whole Brillouin 
zone below the PG temperature, and interestingly seem to connect 
with the ${\bf Q}_{AF}$ resonance observed also in this material.\cite{yu2010}
This again indicates that 
the magnetic properties and the PG behavior observed in charge-like
properties are very likely related.

Out of the sake of completeness, one must mention the spin
response of isolated layers of the parent undoped compound 
La$_2$CuO$_4$, which are only one 
unit cell thick.\cite{dean2012} This response
consisted of the same coherent magnetic excitations, namely magnons,
which occur for the bulk order. Because long-range order cannot exist
in a single layer, the magnons in a single layer need to find
an explanation outside of the conventional spin-wave theory.

Consequently, a theoretical model for the high-$T_C$ 
materials must not only account for the PG and the magnetic resonance 
phenomena separately, but also take into account the feature in iii, 
which suggests that what causes
the PG is perhaps responsible for the magnetic 
resonance as well. This model ought to account also for the magon-like 
response obtained for a single layer
 of La$_2$CuO$_4$, at least qualitatively for now.
In this paper, we argue that
the rotating antiferromagnetism theory (RAFT)
satisfies this criterion.\cite{azzouz2003,azzouz2003p,azzouz2004,azzouz2005,azzouz2010,azzouz2010p,azzouz2012,azzouz2012p,azzouz2013}
RAFT, which was originally proposed in order to explain 
the PG behavior in these materials, yield results in general
consistent with experimental data for 
angle-resolved-photoemission,\cite{azzouz2010} 
optical conductivity,\cite{azzouz2005,azzouz2012}
Raman,\cite{azzouz2010} and thermodynamic properties.\cite{azzouz2003p,azzouz2004}
This theory is based on the phenomenon
of the rotating antiferromagnetic (RAF) order 
whose order parameter is a vector magnetization with a nonzero magnitude 
and a time-dependent phase. This time dependence
makes of the RAF order
an example of hidden order in a spin-liquid state.
RAFT is one among other theoretical 
proposals for 
the PG,\cite{he2011} and is based on spin antiferromagnetism
contrary to the
theories of circulating currents\cite{varma1997,varma2006,chakravarty2001} 
which are based on hidden orbital
antiferromagnetism. Like these theories,
RAFT belongs in the competition 
(between some sort of magnetism and superconductivity) scenario, 
contrary to the theory which is based on the 
preformed superconducting pairs scenario.\cite{emery1995}
An argument in favor of a theory based on spin antiferromagnetism rather than 
one based on orbital antiferromagnetism is that the unusual antiferromagnetic 
${\bf q}=0$ order was observed by counting spin flip events. 
In our opinion, the latter 
can only be defined for a spin $1/2$. If one assumes the conservation 
of the angular momentum in the scattering process, then a spin flip for the neutron
has to be compensated by a spin flip in the excitation of the system, and this can
occur only for the real spin of the electron (or hole); the orbital current 
has an integral angular momentum. If true, this argument will rule out 
the candidacy of any theory 
based on circulating currents for the explanation of the unusual
${\bf q}=0$ antiferromagnetism and the PG. 
We will discuss this point again later in the framework of RAFT.
In the latter, the PG below $T^*$ is caused by
the magnitude of the RAF order parameter. In the present report, 
we argue that the phase of this parameter is responsible
for the magnetic ${\bf Q}_{AF}$ resonance
observed at nonzero doping-dependent frequencies.

The time dependence of the phase of the RAF order parameter 
remained an issue until recently
when a crude estimate was derived in the limit of  
localized electrons, where the effects of the kinetic 
energy and doping were neglected.\cite{azzouz2012p}
This estimate was obtained using the Heisenberg dynamics'
equation where only the onsite Coulomb repulsion
of the Hubbard Hamiltonian was incorporated 
in the equation's commutator, independently of the doping level. 
Note that RAFT is implemented using the two-dimensional Hubbard model.
Even though the physical interpretation of the RAF order 
found within the
crude treatment of Ref. \onlinecite{azzouz2012p} remains correct,
this approximation led to a doping-
and momentum-independent phase. 
%
%The RAF state was interpreted as
%a single $(\pi,\pi)$ spin wave about a zero local magnetization,
%or as a magnetization precessing in the spin's $xy$-plane around zero
%$z$-axis magnetization. 
%
The lack of the doping dependence
made the comparison with available experimental data, 
like those of the neutron scattering experiments, impossible.
It is worth stressing that all the physical properties 
calculated or analyzed so far within RAFT 
do not depend on the phase
of the RAF order parameter. 
Comparison of the results of such works
with experimental data led to satisfactory
agreement in general.
\cite{azzouz2003,azzouz2003p,azzouz2004,azzouz2005,azzouz2010,azzouz2010p,azzouz2012,azzouz2012p,azzouz2013}

The time dependence of the phase of the RAF order parameter
is recalculated in the present work using the total 
RAFT's mean-field Hamiltonian in the Heisenberg equation.
This yields an expression that takes into account
the doping dependence for
the rotational angular frequency $\omega_{sf}$ 
of this order parameter.
The results obtained for $\omega_{sf}$ 
for different Hamiltonian parameters
are discussed in connection with existing 
neutron scattering data.\cite{fujita2012}
The shape of the spin excitations for most of the hole-doped 
materials is the famous hour-glass dispersion, which 
is characterized by an upwardly component separated at the 
waist of the hour glass by a doping dependent 
energy $E_{\rm cross}$ from a downwardly component.
Below $E_{\rm cross}$, the peaks in the magnetic response 
occur at incommensurate momenta, but at $E_{\rm cross}$,
the response's peak occurs at the commensurate momentum $(\pi,\pi)$. 
The rotating magnetization in RAFT introduces 
an energy scale $\hbar \omega_{sf}$, which, as we argue here,
is identified with $E_{\rm cross}$, and is thus
related to the ${\bf Q}_{AF}\equiv(\pi,\pi)$ resonance 
observed in neutron scattering experiments.

The remainder of this paper is organized as follows. First, in order
to estimate the contribution of the RAF order in the neutron scattering
experiments, we analyze the scattering cross-section for this hidden order. 
Then, 
we calculate the time dependence of the phase 
of the rotating order parameter, and explain its connection
with this cross-section.  
Afterward, we calculate the doping dependence of the frequency
of rotation and discuss it in terms of the magnetic resonance
energy measured experimentally.
Finally, conclusions are drawn.

\section{Approach}

\subsection{Calculation of the neutron scattering 
cross-section for RAF order}

To understand the response expected 
from a state with RAF order,
we calculate the contribution of 
such an order to the cross-section starting from 
the well known expression\cite{squires}
\begin{eqnarray}
\frac{d^2\sigma}{d\Omega dE'}&\propto&
%\frac{(\gamma r_o)^2}{2\pi\hbar}\frac{k'}{k}
\sum_{\ell,\ell'}\sum_{\alpha,\beta}
(\delta_{\alpha\beta}-{\hat q}_\alpha{\hat q}_\beta)
%e^{-2W}F^*({\bf q})F({\bf q}) 
e^{i{\bf q}\cdot({\bf R}_{\ell}-{\bf R}_{\ell'})}
\int dt\langle S_{\ell'}^\alpha(0) S_{\ell}^\beta(t)\rangle e^{-i\omega t},
\label{cross section}
\end{eqnarray}
%
%
%where $\gamma=...$ is ..., $r_0=...$, $F({\bf q})$ is the form factor, 
%which we assume constant for convenience.
%
and focus on the spin-spin correlation function. ${\hat q}_{\alpha}$ is 
the {$\alpha$}-component of 
the unit vector ${\hat {\bf q}}={\bf q}/q$; ${\bf q}$ 
is the momentum transfer of the neutron and $\alpha=x, y,z$. 
In RAFT,\cite{azzouz2003}
\begin{eqnarray}
&&\langle S_{\ell}^x\rangle=Q_{\ell}\cos[\phi(t)],\cr
&&\langle S_{\ell}^y\rangle=Q_{\ell}\sin[\phi(t)], \cr
&&\langle S_{\ell}^z\rangle=0,
\label{mag config}
\end{eqnarray}
where $Q_{\ell}=e^{\pm i{\bf Q}_{AF}\cdot{\bf R}_{\ell}}Q$ 
is a staggered magnetization; $Q$
being its magnitude. Here, ${\bf R}_{\ell}=x_{\ell}{\hat x} 
+ y_{\ell}{\hat y}$ designates the coordinates of site $\ell$  
on a two-dimensional lattice. 
$\phi(t)$ is the phase
of the RAF order parameter, whose time 
dependence will be calculated below. But first let us
find out how the cross-section
(\ref{cross section}) depends on 
this phase or its
time derivative (angular frequency). If we were to use
Eqs. (\ref{mag config}) in (\ref{cross section})
for a static helical order with a time-independent phase $\phi$,
then the cross-section would have an elastic component at zero 
energy $\omega=0$.\cite{squires}
For RAF, such an elastic contribution does not exit, but 
a peak 
at a finite frequency (energy) can be shown to exist.

In the limit $t\to\infty$, we use the same approximation that leads
to the elastic contribution for ordinary magnetic 
orders,\cite{squires} namely
$
\lim_{t\to\infty}\langle S_{\ell'}^\alpha(0) S_{\ell}^\beta(t)\rangle
\approx
\langle S_{\ell'}^\alpha(0)\rangle\langle S_{\ell}^\beta(t)\rangle
$.
Normally, for static order, the values of the expectation values
appearing on the right hand side of this equality are the static 
values of the order parameter. Here we replace the time dependence
for $\langle S_{\ell}^\beta(t)\rangle$ by the components of the 
RAF order parameter given by 
expressions 
(\ref{mag config}). As found below,
the consequences 
of this assumption are pretty consistent with the 
neutron-scattering experimental results
regarding the magnetic resonance. 
Calculating
$\langle S_{\ell'}^\alpha(0) S_{\ell}^\beta(t)\rangle$
for $\alpha=x,\ y,\ z$ and $\beta=x,\ y,\ z$ leads to
%
%
%\begin{widetext}
\begin{eqnarray}
&&Q^2\sum_{\ell,\ell'}\int dt
e^{-i\omega t}\big\{\cos[\phi(t)-\phi(0)] -\frac{1}{2}
({\hat q}_x^2 + {\hat q}_y^2)\cos[\phi(t)-\phi(0)]
 \cr 
 && -\frac{1}{2}({\hat q}_x^2 - {\hat q}_y^2)\cos[\phi(t)+\phi(0)]
- {\hat q}_x{\hat q}_y\sin[\phi(t)+\phi(0)\big\}
e^{i({\bf Q}_{AF}+{\bf q})\cdot({\bf R}_\ell-{\bf R}_{\ell'})}.
\label{cross section2}
\end{eqnarray}
%\end{widetext}
%
%
Expression (\ref{cross section2}) becomes 
the same as in the case of a helical arrangement
when $\phi(0)$ is replaced by ${\bf Q}\cdot{\bf R}_\ell$ and $\phi(t)$
by ${\bf Q}\cdot{\bf R}_{\ell'}$.
The vector ${\bf Q}$ is in the direction of the axis of the 
helix, is of magnitude 
$2\pi$ divided by the pitch of the helix,\cite{squires} and should 
not be confused with the RAF order parameter $Q$, which has the physical 
unit of magnetization. As mentioned later on, 
$Q$ is calculated self-consistently 
using RAFT's mean-field equations.\cite{azzouz2003,azzouz2003p,azzouz2004}

Next we Taylor expand $\phi(t)$ to first order in time.
One gets $\phi(t)-\phi(0)\approx\omega_{sf}t$,
where $\omega_{sf}$ is the rotational angular frequency of the RAF order parameter.
Then, $\phi(t)+\phi(0)=\omega_{sf}t +2\phi(0)$. Because there is no reason 
for the initial phase $\phi(0)$ to be non random, averaging over it
must yield
zero for the terms in $\cos[\phi(t)+\phi(0)]$ and $\sin[\phi(t)+\phi(0)]$
in Eq. (\ref{cross section2}), which then reduces to
%
%
%\begin{widetext}
%
\begin{eqnarray}
&&Q^2\sum_{\ell,\ell'}\int dt
e^{-i\omega t}\big[1 -\frac{1}{2}
({\hat q}_x^2 + {\hat q}_y^2)\big]\cos(\omega_{sf}t)
e^{i({\bf Q}_{AF}+{\bf q})\cdot({\bf R_\ell}-{\bf R}_{\ell'})}=\cr
&&
N(2\pi)^2Q^2(1+{\hat q}_z^2)
\sum_{\bf G}\delta({\bf q}+{\bf Q}_{AF}+{\bf G})
\big[ \delta(\omega-\omega_{sf})+\delta(\omega+\omega_{sf})\big],
\label{cross section3}
\end{eqnarray}
%\end{widetext}
%
%
where $N$ is the number of the lattice sites, and 
${\bf G}$ a reciprocal lattice vector. Expression (\ref{cross section3})
is nonzero only when the momentum transfer is 
${\bf q}\equiv {\bf Q}_{AF}$ [mod. a reciprocal lattice vector], and
when the neutron's energy transfer satisfies $\omega=\pm\omega_{sf}$.
The occurrence of a peak in 
Eq. (\ref{cross section3}) at the momentum transfer 
${\bf Q}_{AF}=(\pi,\pi)$ and at a finite energy reminds us 
of the observed magnetic resonance in hole-doped 
high-$T_C$ cuprates.\cite{fujita2012} To push one step 
further the comparison of our results with the experimental data 
for this resonance,
we calculate $\omega_{sf}$ as a function of doping and compare it
with that of the resonance energy or $E_{\rm cross}$.\cite{fujita2012}

\subsection{Calculation of the peak frequency $\omega_{sf}$ within RAFT}

In Ref. \onlinecite{azzouz2012p}, the crude estimate
$\omega_{sf}\approx U/\hbar$ was derived
in the limit of localized electrons, 
where the effects of doping and electrons' 
kinetic energy were neglected.
To get this expression, only the $U$ term of the
Hubbard Hamiltonian was considered in the
Heisenberg equation.
Tremblay,\cite{tremblay2013} however, pointed out 
that the frequency $\omega_{sf}$
needs be calculated using the mean-field RAFT's
Hamiltonian instead of the U-term of the Hubbard model. 
When calculated as such as shown below, 
$\omega_{sf}$ is found to show significant
doping and wavevector dependence.

In order to proceed, we use the definition of the RAF order parameter
$Q_\ell$ in RAFT, namely
$
Q_\ell
%=-\langle S_i^+\rangle
=-\langle c_{\ell\uparrow}^{\alpha\dag} c^{\alpha}_{\ell\downarrow}\rangle
=-\frac{1}{N}\sum_{\bf k} \langle c_{{\bf k}\uparrow}^{\alpha\dag}
c^{\alpha}_{{\bf k}\downarrow}\rangle
$,
with $\ell\in$ sublattice $\alpha=A$ or $B$.
For a site-independent phase, we can get this time dependence by
considering the Green's function
$\langle c_{{\bf k}\uparrow}^{\alpha\dag}(t)
c_{{\bf k}\downarrow}^\alpha(0)\rangle$
with $t>0$. 
Physically, the time dependence of a spin flip process
can be calculated via
$\langle c_{{\bf k}\uparrow}^{A\dag}(t)
c_{{\bf k}\downarrow}^A(0)\rangle$ for sublattice $A$,
which means that a spin-down electron is annihilated 
in state ${\bf k}$ at an earlier time $t=0$, then a spin-up electron is 
created at a later time $t>0$. 
To get the time dependence of
$\langle c_{{\bf k}\uparrow}^{\alpha\dag}(t)
c_{{\bf k}\downarrow}^\alpha(0)\rangle$
we calculate the time dependence of $c_{{\bf k}\uparrow}^{\alpha\dag}(t)$
using the Heisenberg equation with the RAFT Hamiltonian in the commutator.
Note that in Ref. \onlinecite{azzouz2012p}, the time dependence of $S^\pm$
was rather sought. This is one of the reasons why an accurate estimate 
of the frequency was not obtained.

Given the difficulty and 
complexity of the problem under investigation here, we 
restrict our analysis to the  
non superconducting phase. In this case,
RAFT's Hamiltonian is given by\cite{azzouz2013,azzouz2003,azzouz2003p,azzouz2004}
\begin{equation}
H\approx\sum_{{\bf k}\in RBZ}\Psi^{\dag}_{\bf k}{\cal H}\Psi_{\bf k}
+NUQ^2-NUn^2,
\label{raft hamiltonian}
\end{equation}
where 
$Q=|\langle c_{\ell\downarrow}c_{\ell\uparrow}^\dag\rangle|$,
$n=\langle n_{i,\sigma}\rangle$ is the expectation value 
of the number operator, and $U$ is the onsite Coulomb repulsion energy. 
Due to the antiferromagnetic
correlations present in the system even when no 
long-range order occurs well away from half-filling,
the lattice is considered to be made of
two sublattices $A$ and $B$.
The summation runs over the magnetic or reduced
Brillouin zone (RBZ).
The 4-component Nambu spinor is
\begin{eqnarray}
\Psi^\dag=(\Psi_{{{\bf k}}1}^\dag,\Psi_{{{\bf k}}2}^\dag, 
\Psi_{{{\bf k}}3}^\dag,\Psi_{{{\bf k}}4}^\dag)\equiv
(c^{A\dag}_{{{\bf k}}\uparrow},c^{B\dag}_{{{\bf k}}\uparrow},
c^{A\dag}_{{{\bf k}}\downarrow}, c^{B\dag}_{{{\bf k}}\downarrow}),
\label{nambuspinor}
\end{eqnarray}
and the Hamiltonian matrix is 
\begin{eqnarray}
\label{H_density}
{\cal H}=\left(
\begin{array}{cccc}
-\mu'({\bf k})&\epsilon({\bf k})&QU&0\\
\epsilon({\bf k})&-\mu'({\bf k})&0&-QU\\
QU&0&-\mu'({\bf k})&\epsilon({\bf k})\\
0&-QU&\epsilon({\bf k})&-\mu'({\bf k})
\end{array}\right).
\end{eqnarray}
The eigenenergies are $E_{\pm}({\bf k})=-\mu'({\bf k})\pm E_q({\bf k})$, with
$E_q({\bf k})=\sqrt{\epsilon^2({\bf k})+Q^2U^2}$, and
$\mu'=\mu-Un-4t'\cos k_x\cos k_y$; $\mu$ being the chemical potential,
and $\epsilon({\bf k})=-2t(\cos k_x + \cos k_y)$. $t$, ($t'$), is electrons' hopping 
energy between first (second)-nearest neighbors. $Q$ and $n$ are calculated self-consistently
using RAFT's mean-field equations:\cite{azzouz2003}
\begin{eqnarray}
1&=&\frac{U}{2N} \sum_{\bf k} \frac{n_F[E_-({\bf k})]-n_F[E_+({\bf k})]}{E_{q}({\bf k})},\cr
n&=&\frac{1}{2N}\sum_{\bf k}{n_F[E_+({\bf k})]+n_F[E_-({\bf k})]},
\label{mfe}
\end{eqnarray}
where $n_F$ is the Fermi-Dirac factor.
The Hamiltonian (\ref{raft hamiltonian}) can be written in a diagonal form 
using the eigenspinor 
$\Phi^{\dag}_{\bf k}=(\Phi^{\dag}_{{\bf k}1},\Phi^{\dag}_{{\bf k}2},\Phi^{\dag}_{{\bf k}3},
\Phi^{\dag}_{{\bf k}4})$
given by 
$\Phi_{\bf k}=T_{{\bf k}}\Psi_{\bf k}$, where the matrix $T_{{\bf k}}$ is
\begin{eqnarray}
\label{T}
{T_{{\bf k}}}=\frac{1}{\sqrt{2}}\left(
\begin{array}{cccc}
1 &\frac{\epsilon({\bf k})}{E_q({\bf k})} &\frac{QU}{E_q({\bf k})} &  0\\
0 &\frac{QU}{E_q({\bf k})} &-\frac{\epsilon({\bf k})}{E_q({\bf k})} &1 \\
0 &-\frac{QU}{E_q({\bf k})} &\frac{\epsilon({\bf k})}{E_q({\bf k})} &1\\
1 &-\frac{\epsilon({\bf k})}{E_q({\bf k})} &-\frac{QU}{E_q({\bf k})} &0
\end{array}\right).
\end{eqnarray}
%
%
%
%\begin{eqnarray}
%\Phi_{{\bf k}1}&=&\frac{1}{\sqrt{2}}
%\Big\{c^{A}_{{\bf k} \uparrow}+\frac{\epsilon({\bf k})}
%{E_q}c^{B}_{{\bf k} \uparrow}
%+\frac{q}{E_q}c^{A}_{{\bf k} \downarrow}\Big\}\cr
%\Phi_{{\bf k}2}&=&\frac{1}{\sqrt{2}}\Big\{
%\frac{q}{E_q}c^{B}_{{\bf k} \uparrow}
%-\frac{\epsilon({\bf k})}{E_q}c^{A}_{{\bf k} \downarrow}
%+c^{B}_{{\bf k} \downarrow}
%\Big\}\cr
%\Phi_{{\bf k}3}&=&\frac{1}{\sqrt{2}}\Big\{
%-\frac{q}{E_q}c^{B}_{{\bf k} \uparrow}
%+\frac{\epsilon({\bf k})}{E_q}c^{A}_{{\bf k} \downarrow}
%+c^{B}_{{\bf k} \downarrow}
%\Big\}, \cr
%
%
%
%\Phi_{{\bf k}4}&=&\frac{1}{\sqrt{2}}
%\Big\{c^{A}_{{\bf k} \uparrow}-
%\frac{\epsilon({\bf k})}{E_q}c^{B}_{{\bf k} \uparrow}
%-\frac{q}{E_q}c^{A}_{{\bf k} \downarrow}\Big\}.
%\end{eqnarray}
%
%
%
%The Hamiltonian transforms to:
%
%
%\begin{eqnarray}
%H'
%&&=\sum_{{\bf k}\in RBZ}\Phi_{\bf k}^\dag P_{\bf k}{\cal H}
%P_{\bf k}^\dag\Phi_{\bf k} 
%+ NUQ^2-NUn^2 \cr
%&&
%=\sum_{{\bf k}\in RBZ}\Phi^{\dag}_{\bf k}{\cal H}'\Phi_{\bf k}
%+NUQ^2-NUn^2,
%\label{diagonal raft hamiltonian}
%\end{eqnarray}
%
%
%with 
%
%
%
%
%\begin{eqnarray}
%\label{H'_density}
%{\cal H}'=P_{\bf k}{\cal H}P_{\bf k}^\dag=
%\left(
%\begin{array}{cccc}
%E_+({\bf k})&0&0&0\\
%0&E_-({\bf k})&0&0\\
%0&0&E_+({\bf k})&0\\
%0&0&0&E_-({\bf k})
%\end{array}\right).
%\end{eqnarray}
%
%
%
%
%
%
The diagonal Hamiltonian $H'=T_{\bf k}HT_{\bf k}^\dag$ assumes the form
\begin{eqnarray}
&&H'=\sum_{{\bf k}\in RBZ}
\big[E_+(\Phi_{{\bf k}1}^\dag \Phi_{{\bf k}1} 
+\Phi_{{\bf k}3}^\dag \Phi_{{\bf k}3} )
\cr
&&+ E_-(\Phi_{{\bf k}2}^\dag \Phi_{{\bf k}2} +
\Phi_{{\bf k}4}^\dag \Phi_{{\bf k}4})\big]
+ NUQ^2-NUn^2.
\label{diagonal raft hamiltonian2}
\end{eqnarray}
Using the Heisenberg equation, we find
$d\Phi_{{\bf k}j}=\frac{1}{i\hbar}E_\pm({\bf k})\Phi_{{\bf k}j}$ 
($E_+$ for $j=1,3$, and $E_-$ for $j=2,4$), which gives
$\phi_{{\bf k}j}(t)=\phi_{{\bf k}j}(0)e^{-iE_\pm t/\hbar}$. This defines
two characteristic frequencies $\omega_\pm=E_\pm/\hbar$.
We then use
$\Psi_{\bf k}=T^\dag_{{\bf k}}\Phi_{\bf k}$
to calculate 
$c_{{\bf k}\sigma}^{A\dag}(t)$ and $c_{{\bf k}\sigma}^{B\dag}(t)$,
and find
%
%
%\begin{widetext}
\begin{eqnarray}
c^A_{{\bf k}\sigma}(t)&=&\frac{1}{2}\big\{
(e^{-i\omega_+t}+e^{-i\omega_-t})c^A_{{\bf k}\sigma}
+
(e^{-i\omega_+t}-e^{-i\omega_-t})
[\frac{\epsilon}{E_q}c^B_{{\bf k}\sigma}
+
\frac{QU}{E_q}c^A_{{\bf k}{-\sigma}}]
\big\} \cr
c^B_{{\bf k}\sigma}(t)&=&\frac{1}{2}\big\{
(e^{-i\omega_+t}+e^{-i\omega_-t})c^B_{{\bf k}\sigma}
+
(e^{-i\omega_+t}-e^{-i\omega_-t})
[\frac{\epsilon}{E_q}c^A_{{\bf k}\sigma}
-
\frac{QU}{E_q}c^B_{{\bf k}{-\sigma}}]
\big\} 
\label{dcoverdt}
\end{eqnarray}
%\end{widetext}
%
%
with $\sigma=1\equiv\uparrow$ for up spins and $\sigma=-1\equiv\downarrow$
for down ones.

Next, we calculate the expectation values $\langle c^{\alpha\dag}_{{\bf k}\uparrow}(t)
c^{\alpha}_{{\bf k}\downarrow}(0)\rangle$ using the time-independent 
thermal averages $
\langle c^{A(B)\dag}_{{\bf k}\uparrow}
c^{A(B)}_{{\bf k}\downarrow}\rangle=\mp Q$ and
$\langle c^{\alpha\dag}_{{\bf k}\sigma}
c^{\alpha}_{{\bf k}\sigma}\rangle=n$;  $\alpha=A$ or $B$. 
It is found that:
%
%\begin{widetext}
%
\begin{eqnarray}
\langle c^{A(B)\dag}_{{\bf k}\uparrow}(t)
c^{A(B)}_{{\bf k}\downarrow}(0)\rangle=
%-\langle c^{B\dag}_{{\bf k}\uparrow}(t)c^{B}_{{\bf k}\downarrow}(0)\rangle=
\mp\frac{Q}{2}\big[(e^{i\omega_+t} + e^{i\omega_-t}) +
\frac{Un}{E_q}(e^{i\omega_+t} - e^{i\omega_-t})\big],
%
%
%c^{B}_{{\bf k}\downarrow}(0)\rangle&=&
%\frac{1}{2}Q(e^{i\omega_+t} + e^{i\omega_-t}) -
%\frac{QU}{E_q}n(e^{i\omega_+t} - e^{i\omega_-t}).
\label{phase}
\end{eqnarray}
%
%\end{widetext}
%
%
where the $-$, ($+$), corresponds to the $AA$, ($BB$), expectation value.
If we assume, for the sake of simplicity, that the dominant 
contribution comes from the neighborhood of
the Fermi surface, where $E_\pm=\hbar\omega_\pm\sim0$,\cite{azzouz2010,azzouz2013}
then the term in $Un/E_q$ in Eq. (\ref{phase}) can be neglected.
Using 
$e^{i\omega_+t}+e^{i\omega_-t}
=2\cos\big(\frac{\omega_+-\omega_-}{2}t\big)
e^{i\big(\frac{\omega_++\omega_-}{2}t\big)}$,
we get 
\begin{eqnarray}
\langle c^{A,B\dag}_{{\bf k}\uparrow}(t)
c^{A,B}_{{\bf k}\downarrow}(0)\rangle\approx
\mp Qe^{i(\omega_++\omega_-)t/2}.
\label{phase2}
\end{eqnarray}
For the wavevectors
in the immediate neighborhood of the Fermi surface, 
we approximated the cosine term by $1$ in (\ref{phase2}), and
the phase in the complex exponential can be used to define the following
${\bf k}$-dependent spin-flip frequency (let $\hbar=1$):
\begin{eqnarray}
\omega_{sf}({\bf k})=\frac{1}{2}(\omega_++\omega_-)
=-\mu+Un-4t'\cos k_x\cos k_y.
\label{omegasfk} 
\end{eqnarray}
Note that the only ${\bf k}$ dependence of $\omega_{sf}$ appears in
the term in $t'$; $\mu$, $n$, and $U$ being ${\bf k}$
independent.
In the underdoped regime of the p-type cuprates, 
the shape of the Fermi surface resembles elongated ovals that tend to reach out to 
the points $(\pm\pi,0)$ and $(0,\pm\pi)$, and
consists of contours around the points
$(\pm\pi/2,\pm\pi/2)$.\cite{azzouz2013,azzouz2010}
For a given $\mu$ and for a given set of Hamiltonian parameters,
the lowest spin-flip frequency is
realized at the hot spots where the Fermi 
surface intersects the RBZ, and 
not far from the points ${\bf k}=(\pm\pi,0)$ and 
$(0,\pm\pi)$. For n-type underdoped cuprates, the FS is made of pockets around
the points $(\pm\pi,0)$ and $(0,\pm\pi)$.\cite{azzouz2013}
If we let, for simplicity, $k_x=\pi$ and $k_y=0$ Eq. (\ref{omegasfk})
for both types of cuprates, then one gets
\begin{equation}
\omega_{sf}=-\mu+Un+4t'\ \ (t'<0).
\label{omegasf}
\end{equation}
At half-filling, where $n=1/2$, the expression (\ref{omegasf})
becomes half the crude expression derived in Ref. \onlinecite{azzouz2012p}
when we neglect the chemical potential and the kinetic energy term in $t'$.
The doping dependence of the chemical potential $\mu$ 
has already been studied in past 
publications.\cite{azzouz2003,azzouz2003p,azzouz2004}
Because of this doping dependent chemical potential, and the 
relation $p=1-2n$,
the frequency $\omega_{sf}$ shall present significant doping 
dependence as we show below.

\section{Doping dependence of the frequency $\omega_{sf}$}

Eq. (\ref{omegasf}) is the central result of the present work.
We identify the phase of the order parameter defined in 
Eqs. (\ref{mag config}) by writing 
$\phi(t)\approx\omega_{sf}t$ in the vicinity of $(\pi,0)$. This gives
$Q_i\approx\pm|Q_i|e^{i\omega_{sf} t}$;
$+$ and $-$ for site $i\in$ sublattice $B$ and $A$, respectively. 
It is then clear that the cross-section (\ref{cross section3})
will display a peak at $\omega=\pm\omega_{sf}$. 

\begin{figure}
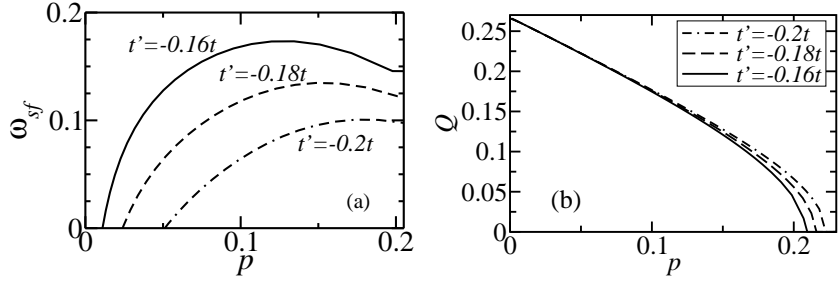

\begin{center}
\includegraphics[height=3.65cm]{./SFfrequency.eps}\ \ 
\includegraphics[height=3.5cm]{./Qvsdoping.eps}
\end{center}
\caption{The spin-flip frequency $\omega_{sf}$ (a) and 
the RAF parameter $Q$ (b) are displayed versus 
doping for $U=2.8t$ and the values of three values of $t'$ as shown.
Temperature is $0.1t$. $n$ and $Q$ are calculated using 
RAFT's mean-field equations (\ref{mfe}).\cite{azzouz2003,azzouz2003p,azzouz2004}
The chemical potential $\mu$ is calculated versus doping $p=1-2n$ 
using the same equations.}
\label{Fig omegasf}
\end{figure}

The spin flips are
purely quantum events, which have been
experimentally measured by Fauqu\'e {\it et al.} using polarized
neutron scattering.\cite{fauque2006} The measurement of these events 
indicated the occurrence of
a new unconventional order below the PG temperature
$T^*$, an order that breaks no 
symmetry.
This order was interpreted by these authors as a zero-momentum (${\bf q}=0$) 
transfer orbital antiferromagnetism 
using the circulating currents' theory.\cite{varma1997,varma2006,chakravarty2001}
However, since the RAF order is induced by the spin-flip processes, we argue
that it is natural to propose that what Fauqu\'e and coworkers observed is rather
rotating antiferromagnetism. The magnitude ($\sim 0.05$ to $0.1$ in units of
$\mu_B$) of the order parameter
they deduced using their measurements
is in good agreement with the values of $Q$ in the underdoped regime,
which satisfy $Q\sim 0.2$ deep in the 
underdoped regime, but drop to about $0.05$ to $0.1$ near
the optimal doping, Fig. (\ref{Fig omegasf}). Moreover, the fact that the order measured 
by Fauqu\'e {\it et al.}
breaks neither translational nor rotational 
symmetry can be attributed to
the rotation of the local magnetization in RAFT; i.e.,
to the time dependence of the phase $\phi\approx\omega_{sf}t$.
We expect that if the system's spins revolve at least once 
while they are being probed by the neutron's spin then no conventional 
order will be detected; 
this means that a net zero local magnetization would be measured
due to the averaging over the phase variations.
We believe that this is the reason why no magnetic order has been measured 
using non polarized neutrons. Polarized neutrons can however 
be used to count the spin flip events rather than
measuring the magnitude of the magnetization directly like in the case 
of a conventional order.

In RAFT, the definition of $Q_i=\langle c_{i\uparrow}c_{i\downarrow}^\dag\rangle$
indicates that the magnitude $Q=|Q_i|$ can be interpreted 
as the probability for a spin-flip event to occur. 
This probability can also be written macroscopically as
the ratio of the number of the spin-flip events over the total number of events, 
which is the sum of the number of the spin-flip events and non spin-flip events in 
a given experiment that is capable of counting such events. We believe that 
this is what the polarized
neutron scattering experiment of Fauqu\'e {\it et al.} did.

%
%{\bf 
%Discuss the observation of spin-wave type excitations in a single CuO$_2$
%layer, and the fact that this response is very similar to that 
%of LaCuO$_4$??????
%
%When the fluctuations around the phase difference $\phi_{i}-\phi_{j}=\pi$ (where $i$
%and $j$ are any two adjacent sites) are considered, the RAFT Hamiltonian 
%becomes complex because no gauge transformation can be used to absorb the phase terms
%in the Hamiltonian. This implies that time reversal symmetry will be broken.
%
%...
%
%}
%
%

We now discuss the results of the calculation of $\omega_{sf}$ as a function of doping
for the Hamiltonian parameters $U=2.8t$ with three different values of $t'$;
$t'=-0.16t$, $-0.18t$, and $-0.2t$. The temperature is $T=0.1t$.
Fig. \ref{Fig omegasf} displays the results of such calculations.
The mean-field equations (\ref{mfe}) were solved numerically using a C code.
It is very interesting that $\omega_{sf}$ shows a doping dependence
similar to that of the energy $E_{\rm cross}$
defined as the energy at the waist of the hour-glass
spectrum displayed by the high-$T_C$ superconducting 
families La$_{2-x}$Sr$_x$CuO$_4$, 
La$_{2-x}$Ba$_x$CuO$_4$, YBa$_2$Cu$_3$O$_{6+x}$, 
and Bi$_2$Sr$_2$Cu$_2$O$_{8+\delta}$ 
(refer to Fig. 2 of Ref. \onlinecite{fujita2012} for the hour-glass spectra
and to Fig. 5 of the same reference for the 
doping dependence of $E_{\rm cross}$).
In the underdoped regime, both $E_{\rm cross}$ and $\omega_{sf}$
increase with doping, reach a maximum at a doping below 
the optimal point, then decrease slightly. Note that $\omega_{sf}$
is meaningful only below the optimal doping because 
the RAF order disappears at a quantum critical point which coincides  practically with 
this point.\cite{azzouz2003}
This result is not in contradiction with some of the experimental results
which also indicate that the resonance happens only in the 
uderdoped regime.\cite{fujita2012}

\section{Discussion}

Together with the broken time-reversal symmetry observed 
in photoemission experiments,\cite{kaminski2002} polarized neutron 
diffraction experiments\cite{fauque2006,li2008} indicated the
universal existence of the ${\bf q}=0$ unusual magnetic order below $T^*$.
In addition, the inelastic neutron scattering data reported 
for HgBa$_2$CuO$_{4+\delta}$ by Li {\it et al.}\cite{li2010}
revealed a fundamental collective magnetic mode
associated with this order. This collective 
mode seems to connect with the
magnetic resonance that occurs at ${\bf Q}_{AF}=(\pi,\pi)$.
If this connection is confirmed 
then the zero-momentum (${\bf q}=0$) order and the 
${\bf Q}_{AF}$ resonance will have to be interpreted 
as two different signatures
for the same physical phenomenon, 
which we propose here to be the 
rotating antiferromagnetic order.
This suggests that the decoration of a unit cell
with an even number of magnetic moments, like in Varma's 
theory,\cite{varma1997,varma2006}
would not be adequate, as this would rule out 
any staggered antiferromagnetic correlation at ${\bf Q}_{AF}$.

This claim is supported in our opinion by other
experimental data collected so far within the PG, 
like the spin-like excitations reported in the single layer
Bi$_{2+x}$Sr$_{2-x}$CuO$_{6+\delta}$,\cite{enoki2010}
the static or quasi-static incommensurate
spin order observed at low temperature 
in YBa$_2$Cu$_3$O$_{6+x}$ with $x = 0.45$ and $2\%$ Zn-doped
$x = 0.6$ crystals,\cite{hinkov2008,suchaneck2010}
or the hour-glass spin excitations\cite{fujita2012} reported 
in 
La$_{2-x}$Sr$_x$CuO$_4$, 
La$_{2-x}$Ba$_x$CuO$_4$, YBa$_2$Cu$_3$O$_{6+x}$, 
and Bi$_2$Sr$_2$Cu$_2$O$_{8+\delta}$
systems. All these results are interpreted in terms of 
excitations of the spin 
degrees of freedom.

Remnant magnetic excitations appear also to survive
at higher energies in the underdoped phase. Above 
$E_{\rm cross}$ in the hour-glass spectrum, 
these excitations resemble
the spin wave excitations in the 
undoped cuprate parents. Below $E_{\rm cross}$,
the correlations lead however to new
magnetic excitations. These results constitute also
significant evidence for the nature and origin 
of these excitations being the spin degrees of freedom.
All the above discussion and the results
of the present work suggest that RAFT, which 
is based on spin antiferromagnetism,
is a serious candidate for modeling the unusual 
magnetic properties. This adds to the fact that RAFT
has been successfully used to model the electronic
properties of the high-$T_C$ cuprates.

\section{conclusion}
\label{sec3}

In summary, the neutron-scattering response expected from 
a state characterized 
by rotating antiferromagnetic order is calculated in this paper 
within the rotating antiferromagnetism theory. We argue 
that the resonance peak observed experimentally in the hole-doped
high-$T_C$ 
cuprate superconductors is a consequence of
the rotating antiferromagnetic order. 
We find that the phase of the order parameter of this order
is responsible for the occurrence of a peak at a nonzero frequency,
for which we estimated its doping dependence. The trends shown
by this dependence are very similar to those of the
doping dependence of the resonance energy
or the energy $E_{\rm cross}$ at the waist of the
magnetic excitations spectrum, namely the hour-glass spectra.
The order parameter of the rotating antiferromagnetic order
has a magnitude and a phase.\cite{azzouz2003} 
From this work and earlier ones,\cite{azzouz2013}
it turns out that while
the magnitude is responsible for the pseudogap behavior and other 
unusual thermodynamic and transport properties, the phase
is responsible for the unusual magnetic properties
of the high-$T_C$ cuprate materials, at least in the hole doped materials.
If our present claim of RAFT being able to
account for the unusual magnetic properties
is confirmed by other independent works,
then we will be one more step closer to the applicability
of the rotating antiferromagnetism theory for the high-$T_C$ materials.

\acknowledgments{
The author would like to thank 
Andr\'e-Marie Tremblay for his critical reading of the manuscript and his comments
}
%\end{acknowledgements}


\begin{thebibliography}{00}

\bibitem{fauque2006} B. Fauqu\'e, Y. Sidis, V. Hinkov, S. Pailh\`es, C.T. Lin, 
X. Chaud, and P. Bourges, Phys. Rev. Lett. {\bf 96}, 197001 (2006).

\bibitem{li2008} Y. Li, V. Bal\'edent, N. Barisic, Y. Cho, 
B. Fauqu\'e, Y. Sidis, G. Yu, X. Zhao, P. Bourges, and M. Greven,
Nature {\bf 455}, 375 (2008).


\bibitem{timusk1998} T. Timusk, B. Statt, Rep. Prog. 
Phys. {\bf 62}, 61 (1999).

\bibitem{fujita2012} For a review on the unusual magnetic properties of 
HTSCs refer to:
Masaki Fujita, Haruhiro Hiraka, Masaaki Matsuda,
Masato Matsuura, John M. Tranquada, Shuichi Wakimoto, Guangyong Xu,
Kazuyoshi Yamada, 
J. Phys. Soc. Jpn. {\bf 81}, 011007 (2012).

\bibitem{rossat-mignod1991} J. Rossat-Mignod, L.P. Regnault, C. Vettier,
P. Bourges, P. Burlet, J. Bossy, 
J. Y. Henry, and G. Lapertot,
Physica C {\bf 185-189}, 86 (1991).

\bibitem{li2010} Y. Li, V. Bal\'dent, G. Yu, N. Barisic, K. Hradil, 
R.A. Mole, Y. Sidis, P. Steffens, X. Zhao, P. Bourges, M. Greven,
Nature {\bf 468}, 283 (2010).

\bibitem{yu2010} G. Yu, Y. Li, E. M. Motoyama, X. Zhao, N. Barisic, Y. Cho, P. Bourges, 
K. Hradil, R. A. Mole, and M. Greven, Phys. Rev. B {\bf 81}, 064518 (2010).


\bibitem{dean2012} M. P. M. Dean, R. S. Springell, C. Monney, K. J. Zhou, 
J. Pereiro, I. Bozovic, B. Dalla Piazza,
H. M. Ronnow, E. Morenzoni, J. van den Brink, T. Schmitt, and J. P. Hill,
Nature Materials {\bf 11}, 850 (2012).


\bibitem{azzouz2003}M. Azzouz, Phys. Rev. B {\bf 67}, 134510 (2003).


\bibitem{azzouz2003p}M. Azzouz, Phys. Rev. B {\bf 68}, 174523  (2003).

\bibitem{azzouz2004}M. Azzouz, Phys. Rev. B {\bf 70}, 052501  (2004).

\bibitem{azzouz2005}H. Saadaoui, M. Azzouz, 
Phys. Rev. B {\bf 72}, 184518 (2005).


\bibitem{azzouz2010} M. Azzouz, K.C. Hewitt, H. Saadaoui, 
Phys. Rev. B
{\bf 81}, 174502 (2010).

\bibitem{azzouz2010p} M. Azzouz, B.W. Ramakko, G. Presenza-Pitman,
J. Phys.: Condens. Matter {\bf 22}, 345605 (2010).

\bibitem{azzouz2012} E.H. Bhuiyan, G. Presenza-Pitman, M. Azzouz,
Physica C {\bf 473}, 61 (2012).

\bibitem{azzouz2012p} M. Azzouz, Physica C {\bf 480}, 34  (2012).

\bibitem{azzouz2013}M. Azzouz, Spectrum {\bf 5}, 215 (2013).

\bibitem{he2011} Rui-Hua He, M. Hashimoto, H. Karapetyan, J.D. Koralek, 
J.P. Hinton, J.P. Testaud, V. Nathan, Y. Yoshida, Hong Yao, 
K. Tanaka, W. Meevasana, R.G. Moore, D. H. Lu,
S.-K. Mo, M. Ishikado, H. Eisaki, Z. Hussain, T.P. Devereaux,
S.A. Kivelson, J. Orenstein, A. Kapitulnik, Z.-X. Shen,
Science {\bf 331}, 1579 (2011).


\bibitem{varma1997} C.M. Varma, Phys. Rev. B {\bf 55}, 14554 (1997).

\bibitem{varma2006} C.M. Varma, Phys. Rev. B {\bf 73}, 155113 (2006).

\bibitem{chakravarty2001} S. Chakravarty, R.B. Laughlin, D.K. Morr, 
C. Nayak, Phys. Rev. B {\bf 63},  094503 (2001).

\bibitem{emery1995} V.J. Emery and S.A. Kivelson, Nature {\bf 374}, 434 (1995).


\bibitem{squires} For all the quantities involved in this expression 
refer to: G.L. Squires, Introduction to the Thermal 
Neutron Scattering, Dover Publications, INC. 1978.

\bibitem{tremblay2013} Private email communication (2013).

\bibitem{enoki2010} M. Enoki, M. Fujita, S. Iikubo, and K. Yamada, Physica C
{\bf 470}, S37 (2010).

\bibitem{kaminski2002} A. Kaminski,
S. Rosenkranz, H. M. Fretwell, J. C. Campuzano, Z. Li, H. Raffy, 
W. G. Cullen, H. You, C. G. Olson, C. M. Varma, and H. Hochst,
Nature {\bf 416}, 610 (2002).

\bibitem{hinkov2008} V. Hinkov, D. Haug, B. Fauqu\'e, P. Bourges, Y. Sidis,
A. Ivanov, C. Bernhard, C. T. Lin, and B. Keimer, Science
{\bf 319}, 597 (2008).

\bibitem{suchaneck2010} A. Suchaneck, V. Hinkov, D. Haug, L. Schulz, C. Bernhard,
A. Ivanov, K. Hradil, C. T. Lin, P. Bourges, B. Keimer, and
Y. Sidis, Phys. Rev. Lett. {\bf 105}, 037207 (2010).


\end{thebibliography}
\end{document}